\documentclass[a4paper,fleqn]{cas-dc}
\usepackage[numbers,sort&compress]{natbib}

\makeatletter
  \DeclareSymbolFont{cmsymbols}{OMS}{cmsy}{m}{n}
  \SetSymbolFont   {cmsymbols}{bold}{OMS}{cmsy}{b}{n}

  \DeclareMathSymbol{\cdot} {\mathbin}{cmsymbols}{"01} % ·
  \DeclareMathSymbol{\times}{\mathbin}{cmsymbols}{"02} % ×
  \DeclareMathSymbol{\pm}    {\mathbin}{cmsymbols}{"06} % ±
  \DeclareMathSymbol{\mp}    {\mathbin}{cmsymbols}{"07} % ∓
\makeatother

\usepackage{stfloats}
\usepackage{float}
\restylefloat{table}
\usepackage{adjustbox}
\usepackage{caption}
\usepackage{subcaption}
\usepackage{hyperref}
\usepackage{color}
\usepackage{tikz}
\usepackage{multirow}%

\usepackage{textcomp}%
\usepackage{manyfoot}%
\usepackage{booktabs}%
\usepackage{listings}%
\usepackage{rotating}
\usepackage{placeins}
\usepackage{graphicx}

\usepackage{lmodern} 
\usepackage[absolute,overlay]{textpos}
\usepackage{svg}
\usepackage{enumerate}

\usepackage{lmodern}

\usepackage{tabularx, makecell, booktabs} % in the preamble
\newcolumntype{C}{>{\centering\arraybackslash}X} % centred X

\def\tsc#1{\csdef{#1}{\textsc{\lowercase{#1}}\xspace}}
\tsc{WGM}
\tsc{QE}
\begin{document}
\let\WriteBookmarks\relax
\def\floatpagepagefraction{1}
\def\textpagefraction{.001}

\shorttitle{}    

\shortauthors{}  

\title [mode = title]{Benchmarking Ophthalmology Foundation Models for Clinically Significant Age Macular Degeneration Detection}  



%%%%%%%%%%%%%%%%%%%


\author[1,2]{Benjamin A. Cohen}%[<options>]


%\ead{}


\credit{Developed the algorithms and performed the analysis under the supervision of JAB, drafted the first version of the manuscript}

\affiliation[1]{organization={Faculty of Biomedical Engineering, Technion-IIT},
            city={Haifa},
            country={Israel}}

\affiliation[2]{organization={Taub Faculty of Computer Science, Technion-IIT},
            city={Haifa},
            country={Israel}}

%%%%%%%%%%%%%%%%%%%%%

\author[1,3]{Jonathan Fhima}%[]

\credit{Assisted BAC in the technical development}

\affiliation[3]{organization={Faculty of Mathematics, Technion-IIT},
            city={Haifa},
            country={Israel}}

%%%%%%%%%%%%%%%%%%%%%%

\author[4,5]{Meishar Meisel}%[]

\credit{Curated the HYAMD dataset, provided medical guidance}

\affiliation[4]{organization={Department of Ophthalmology, Hillel Yaffe Medical Center},
            city={Hadera},
            country={Israel}}

\affiliation[5]{organization={The Ruth and Rappaport Faculty of Medicine, Technion-IIT},
            city={Haifa},
            country={Haifa}}

%%%%%%%%%%%%%%%%%%%%%%

\author[4,5]{Baskin Meital}%[]

\credit{Curated the HYAMD dataset}

%%%%%%%%%%%%%%%%%%%%

\author[6,7]{Luis F. Nakayama}%[]

\credit{Contributed and curated the BRAMD dataset, provided medical guidance}

\affiliation[6]{organization={Ophthalmology Department, São Paulo Federal University},
            city={São Paulo},
            country={Brazil}}

\affiliation[7]{organization={Medical Engineering and Science, Massachusetts Institute of Technology},
            city={Cambridge},
            country={USA}}

%%%%%%%%%%%%%%%%%%%%%

\author[4,5,8]{Eran Berkowitz}%[]

\credit{Provided medical guidance throughout the study, contributed and curated the HYAMD dataset}

\affiliation[8]{organization={The Adelson School of Medicine, Ariel University},
            city={Ariel},
            country={Israel}}
            
%%%%%%%%%%%%%%%%%%%%%

\author[1]{Joachim A. Behar}%[]

%\cormark[1]

\credit{Conceived and designed the research, drafted the first version of the manuscript. We acknowledge the assistance of ChatGPT, an AI-based language model developed by OpenAI, in editing the manuscript}

%%%%%%%%%%%%%%%%%%%%%


\begin{abstract}
Self-supervised learning (SSL) has enabled Vision Transformers (ViTs) to learn robust representations from large-scale natural image datasets, enhancing their generalization across domains. In retinal imaging, foundation models pretrained on either natural or ophthalmic data have shown promise, but the benefits of in-domain pretraining remain uncertain. To investigate this, we benchmark six SSL-pretrained ViTs on seven digital fundus image (DFI) datasets totaling 70,000 expert-annotated images for the task of moderate-to-late age-related macular degeneration (AMD) identification. Our results show that iBOT pretrained on natural images, achieves the highest out-of-distribution generalization, with AUROCs of 0.80–0.97, outperforming domain-specific models, which achieved AUROCs of 0.78–0.96 and a baseline ViT-L with no pretraining, which achieved AUROC of 0.68-0.91. These findings highlight the value of foundation models in improving AMD identification, and challenge the assumption that in-domain pretraining is necessary. Furthermore, we release BRAMD, an open-access dataset (n=587) of DFIs with AMD labels from Brazil.
\end{abstract}


\begin{highlights}
\item \textbf{Comprehensive Benchmarking:} We conduct an extensive benchmark across both general and in-domain foundation models performance for the downstream task of intermediate-to-late AMD identification from DFIs.
\item \textbf{AMDNet Model:} Our resulting best model, denoted AMDNet is made available via the Lirot.ai platform.
\item \textbf{New Open-Access Dataset:} We introduce and publicly release BRAMD (n=587), a new DFI dataset with expert-annotated labels for AMD.
\end{highlights}

%\nocite{*}

% Keywords

\maketitle

% Main text

%%%%%% Dataset Table %%%%%%
\begin{table*}[t]
\caption{Description of the study datasets, including the newly introduced BRAMD dataset.}
\centering
\label{tab:datasets}
\resizebox{2\columnwidth}{!}{%
\renewcommand{\arraystretch}{2.2}
\begin{tabular}{cccccccccc}
\textbf{Dataset} & \textbf{DFI} & \textbf{Patients} & \textbf{Device} & \textbf{FOV} & \textbf{Resolution} & \textbf{AMD\%} & \textbf{Sex (m\%)} & \textbf{Age} & \textbf{Origin} \\ \hline \hline

AREDS &
  62,769 &
  4,695 &
  Zeiss FF-series &
  30° &
  3400×2300 &
  30 &
  43 &
  55-81 &
  US \\ \hline
RFMiD1 &
  3,200 &
  NA &
  VAR &
  45° / 50° &
  2144×1424 &
  5.2 &
  - &
  - &
  India \\ \hline
HYAMD &
  1,570 &
  433 &
  Topcon &
  45° &
  1960×1934 &
  27 &
  51 &
  57-95 &
  Israel \\ \hline
ADAM &
  1,200 &
  NA &
  \renewcommand{\arraystretch}{1.3}\begin{tabular}[c]{@{}c@{}}Zeiss Visucam 500 \\ CanonCR-2\end{tabular} &
  45° &
  \renewcommand{\arraystretch}{1.3}\begin{tabular}[c]{@{}c@{}}2124×2056  \\  1444×1444\end{tabular} &
  19 &
  53 &
  37-69 &
  China \\ \hline
FIVES &
  800 &
  573 &
  TRC-NW8 &
  50° &
  2048×2048 &
  25 &
  41 &
  18-80 &
  China \\ \hline
\renewcommand{\arraystretch}{1.2} \begin{tabular}[c]{@{}c@{}} BRAMD \\ {\scriptsize \textit{(new contribution)}} \end{tabular} &
  587 &
  472 &
  \renewcommand{\arraystretch}{1.4}\begin{tabular}[c]{@{}c@{}}CANON CR \\ NIKON NF5050\end{tabular} &
  50° &
  \renewcommand{\arraystretch}{1.4}\begin{tabular}[c]{@{}c@{}}2390×1880 \\  2672×2056\end{tabular} &
  50 &
  37 &
  18-94 &
  Brazil \\ \hline
STARE &
  397 &
  NA &
  Topcon &
  35° &
  700×605 &
  11 &
  - &
  - &
  US \\ %\hline
\end{tabular}%
}
\end{table*}
%%%%%%%%%%%%

\section{Introduction}\label{intro}

Deep learning has become a reliable and practical tool for the automated identification of ophthalmic and systemic diseases from retinal scans \cite{Poplin2018, Ting2019, Cheung2022, Srivastava2023, 
 Men2025, abramovich2025gonetgeneralizabledeeplearning}. Among existing machine learning methods, self-supervised learning (SSL) has emerged as a powerful approach for pretraining Vision Transformers (ViTs) on large-scale image datasets, yielding strong feature representations that generalize across diverse domains. SSL enables models to extract meaningful latent representations from unlabeled data, bypassing the need for explicit annotations. This approach leverages vast quantities of unannotated images to create foundation models pretrained on large datasets, which can then be efficiently fine-tuned for downstream tasks with limited labeled data. In medical imaging, SSL-pretrained models have demonstrated significant performance gains in diagnostic applications \cite{Krishnan2022}. In retinal image analysis, foundation models pretrained on either natural images or in-domain ophthalmic data have demonstrated improved performance across various tasks \cite{retfound2023, visionfm2024, shi2024eyefoundmultimodalgeneralistfoundation}. However, it remains unclear to what extent in-domain pre-training provides an advantage over training on natural images. Indeed, while recent retinal foundation models \cite{retfound2023, visionfm2024, shi2024eyefoundmultimodalgeneralistfoundation} have claimed improvement over foundation models pretrained on natural images, our recent observations challenge this claim \cite{Men2025, abramovich2025gonetgeneralizabledeeplearning}. This raises the need for a deeper investigation of the matter. We propose investigating this question by benchmarking modern SSL-pretrained ViTs on either natural or ophthalmic images for the task of identifying intermediate-to-late age-related macular degeneration (AMD) from retinal scans. 

%with the global prevalence expected to rise from 200 million today to 288 million by 2040 \cite{WONG2014e106}

AMD is a leading cause of irreversible vision loss in adults over 40, with a prevalence of 12.6\% in the US in 2019, projected to affect 288 million people globally by 2040 \cite{WONG2014e106, USprevalence2022}. It impairs central vision, significantly impacting daily tasks such as reading, driving, and facial recognition. Current diagnostic methods rely heavily on specialized ophthalmic examinations, limiting accessibility. Computer-aided analysis of digital fundus images (DFIs) offers a scalable and non-invasive alternative for AMD detection. In this study, we specifically focus on identifying clinically significant AMD, defined as moderate to late stages, as these are the stages associated with significant risk of vision loss.

To conduct our experiments, we utilize seven DFI datasets comprising 70,000 images and benchmark six foundation models \cite{mae2021, Mugs2022, zhou2022ibotimagebertpretraining, retfound2023, dinov2-2024, visionfm2024}, emphasizing out-of-domain (OOD) generalization. OOD evaluation assesses the robustness of models to variations in geography, imaging conditions, and devices, providing a more realistic measure of clinical applicability compared to traditional external dataset fine-tuning. This research makes three key scientific contributions:

\begin{itemize}
\item \textbf{Comprehensive Benchmarking:} We conduct an extensive benchmark across both general and in-domain foundation models performance for the downstream task of intermediate-to-late AMD identification from DFIs.
\item \textbf{AMDNet Model:} Our resulting best model, denoted AMDNet is made available via the Lirot.ai platform.
\item \textbf{New Open-Access Dataset:} We introduce and publicly release BRAMD (n=587), a new DFI dataset with expert-annotated labels for AMD.
\end{itemize}

%%%%%%%%%%%%%%%%%%%%%%%%%%
\section{Datasets}\label{datasets}
We curated a comprehensive collection of 70,523 DFIs from seven independent datasets that encompass diverse imaging conditions and field of view (FOV). This collection includes six publicly available datasets and one newly introduced dataset. Table \ref{tab:datasets} provides a detailed summary of the technical and demographic characteristics of these datasets. For all datasets, we classified images into two categories: intermediate-to-late AMD (denoted as "AMD") and non-intermediate-to-late AMD (denoted as "non-AMD"). Specifically, we defined intermediate-to-late AMD as eyes diagnosed with wet AMD, also referred to as neovascular AMD (NVAMD), visible geographic atrophy (GA), or with the presence of large drusen (LD) \cite{Ferris2005}.

%%%%%%%%%
\textbf{BRAMD:} Developed in collaboration with São Paulo Federal University (São Paulo, Brazil; UNIFESP IRB approval No. 49171021.6.0000.5505), BRAMD includes 587 DFIs from 472 patients. The AMD cohort (295 images) comprises patients aged 54–94, with 237 cases of wet AMD. The control group includes 292 images from diabetic retinopathy patients aged 18–91. Diagnoses were confirmed through clinical examinations supported by OCT scans. Images were acquired using two devices: Canon CR (437 images at 2390\texttimes{}1880 resolution) or Nikon NF5050 (158 images at 2672\texttimes{}2056 resolution). BRAMD has been made open-access on PhysioNet \cite{physionet2000} as a contribution of this research work.

%%%%%% AREDS classification table %%%%%%

\begin{table}[b]

\centering
\renewcommand{\arraystretch}{1.4} % Increases row height for better spacing
\begin{tabular}{p{5cm}|c}
\textbf{Category} & \textbf{Number of Images} \\ \hline \hline
\textbf{AMD} & \textbf{35,668} \\ 
\hspace{1em} NVAMD & 12,945 \\ 
\hspace{1em} GA & 6,158 \\ 
\hspace{1em} Both NVAMD and GA & 3,403 \\ 
\hspace{1em} Large Drusen & 13,162 \\ \hline
\textbf{Non-AMD} & \textbf{27,101} \\ 

\hspace{1em} Other non Control & 14,280 \\ 
\hspace{1em} Control & 5,882 \\ 
\hspace{1em} Control Questionable & 6,939 \\ \hline
\textbf{Excluded categories} & \textbf{11,629} \\ 
\hspace{1em} Questionable AMD & 3,386 \\ 
\hspace{1em} Large Drusen Questionnable & 8,243 \\ \hline
\textbf{Total} & \textbf{74,398} \\  \hline
\end{tabular}
\caption{Distribution of AREDS Classification of Images}
\label{tab:areds_classification}

\end{table}
%%%%%%%%%%%%

\textbf{AREDS} \cite{areds} is one of the largest publicly available AMD fundus image datasets, comprising over 200,000 DFIs from 4,757 patients aged 55–80 across 12 eye clinics in the US. Among available fields, we selected macula-centered color stereo images, which is the most relevant for studying AMD, totaling 140,000 images. In stereo imaging, two images are taken per eye during each visit. We thus selected one image per eye based on predefined quality criteria \cite{Lyu2022}. Patients were categorized based on AMD severity, including those without AMD, and were followed for a median of 6.5 years to monitor disease progression. Diagnoses were determined through fundus photograph grading by a central reading center and ophthalmologic evaluations. Table \ref{tab:areds_classification} summarizes how the different image categories provided by AREDS were labeled as AMD or non-AMD. Categories marked as "Questionable AMD" and "Large Drusen Questionable" were excluded due to uncertainty regarding the eye label. We are thus left with 62,769 images meeting our exclusion criteria. The dataset was split into AREDS-train (90\%) and AREDS-test (10\%), ensuring that images from the same patient did not appear in both subsets to prevent information leakage.

%%%%%%%%%
\textbf{RFMiD1 \cite{s3g7-st65-20}:} is a diverse dataset containing 3,200 DFIs from patients at Sushrusha Hospital, India. Diagnoses were based on fundus photograph labeling and clinical records under expert supervision but did not include OCT imaging. AMD cases include multiple drusen in the macula, GA, or NVAMD. This dataset includes forty-five different eye conditions in total.

\textbf{HYAMD \cite{hyamd2025}:} This recent open dataset was created by Hillel Yaffe Medical Center (Hadera, Israel; Helsinki approval number 0029-24-HYMC), HYAMD contains 1,570 DFIs from 433 patients, with follow-up data spanning 2021 to 2024. The AMD cohort includes 188 patients (95 males, 93 females) aged 57–95, while the control group comprises 192 diabetic retinopathy patients (113 males, 79 females) aged 24–92. DFIs in the AMD cohort either present with NVAMD, GA or LD. Diagnoses were based on comprehensive clinical examinations supported by OCT scans, including OCT angiography and fluorescein angiography when necessary.

\textbf{ADAM \cite{dt4f-rt59-20}} includes 1,200 DFIs from Zhongshan Ophthalmic Center, China. Diagnoses were based on medical records, including retinal examinations (DFI and OCT). AMD cases cover early to advanced stages. To align with our task, we select only intermediate-to-late cases to be part of the positive class while leaving all other cases in the control cohort. 

\textbf{FIVES \cite{fives}} consists of 800 high-quality DFIs collected between 2016 and 2021 at Zhejiang University, China. Diagnoses were based on comprehensive ocular examinations, including slit-lamp examinations, fundus photography, OCT, and fluorescein angiography when necessary. AMD cases present drusen, exudate, geographic atrophy, or hemorrhage, mostly corresponding to intermediate-to-late stages of the disease.

\textbf{STARE \cite{stare}} is a dataset comprising 397 fundus images from the Shiley Eye Center and Veterans Administration Medical Center in San Diego, US. The diagnoses were based on image features rather than comprehensive ophthalmic evaluations. The AMD cases include LD, GA and NVAMD.

%%%%%%%%%%%%%%%%%%%%%%%
\section{Methods}

We benchmarked six state-of-the-art (SOTA) deep learning models fine-tuned on AREDS-train, the largest publicly available dataset for AMD detection. The best performing foundation model is further trained using a multi-source domain training approach, resulting in a final model we call AMDNet.

\subsection*{Data pre-processing}
DFIs are first cropped to remove unnecessary background, then padded to form a square layout to maintain the aspect ratio during resizing. Finally, they are resized to 518×518 pixels using bilinear interpolation \cite{ParsaniaScaling}. This process is used to standardize images from various sources with differing resolutions.

%%%%%% ViT details table %%%%%%
\begin{table}[t]
    \centering
    \renewcommand{\arraystretch}{1.3}
    \setlength{\tabcolsep}{5pt} % Adjust column spacing
    \caption{Technical summary of benchmarked ViT backbones.} %DINOv2 \cite{dinov2-2024} uses a mixed pre-training resolution of 224 and 518.
    \label{tab:ssl_backbones}
\resizebox{\columnwidth}{!}{%
\begin{tabular}{l|c|c|c|c}
\textbf{Model}    & \textbf{Arch.} & \textbf{Params} & \textbf{Pre-training data} & \textbf{\begin{tabular}[c]{@{}c@{}}Pre-training\\ resolution\end{tabular}} \\ \hline
\textbf{MAE} \cite{mae2021}& 
ViT-L/16 & 
307M& 
IN-1k & 
224 \\
\textbf{Mugs} \cite{Mugs2022} & 
ViT-L/16 & 
304M &
IN-1k & 
224\\
\textbf{iBOT} \cite{zhou2022ibotimagebertpretraining}&
ViT-L/16 &
307M &
IN-1k + IN-22k & 
224 \\
\textbf{Dinov2} \cite{dinov2-2024}&
ViT-L/16 &
307M & 
IN-1k + IN-22k & 
224-518 \\
\textbf{RETFound} \cite{retfound2023} & ViT-L/16 &
303M &
IN-1k + 900k DFIs & 
224\\
\textbf{VisionFM} \cite{visionfm2024} & ViT-B/16 &
86M &
IN-1k + 1M DFIs & 
256                                                                       
\end{tabular}}

\end{table}
%%%%%%%%%%%%

%%%
\subsection*{Foundation models}
\label{sec:model_benchmark}
The following ViTs pretrained using SSL on natural images were benchmarked: MAE \cite{mae2021}, Mugs \cite{Mugs2022}, iBOT \cite{zhou2022ibotimagebertpretraining}, and DINOv2 \cite{dinov2-2024}. Additionally, the following ViTs pretrained using SSL on retinal images were evaluated: RETFound \cite{retfound2023} and VisionFM \cite{visionfm2024}. respectively pretrained on  904,170 DFIs and 1,010,293 DFIs. A detailed description of all models is provided in Table \ref{tab:ssl_backbones}. All models were fine-tuned on AREDS-train for 10 epochs with a batch size of 20 DFIs, using a learning rate ranging from 1e-5 to 3e-4 (with 1,000 warmup steps), weight decay of 0.05, and a cosine learning rate scheduler. Data augmentation included random horizontal flips and adjustments to contrast, saturation, and hue. In addition to the six foundation models, we include a baseline ViT-L \cite{dosovitskiy2021imageworth16x16words} with standard initialization and trained with a learning rate of 1e-4, for 60 epochs. For all models, training and evaluation were conducted on NVIDIA A100 GPUs (40GB memory). We evaluate the average AUROC performance of each foundation model and the baseline ViT-L over all datasets, including in-domain performance on AREDS-test and OOD generalization across all the other datasets, in order to identify the best performing foundation model. This foundation model is then selected for multi-source domain training.

%%%
\subsection*{Multi-source domain training}
The best performing foundation model is further trained and evaluated following a methodology established in our previous work \cite{Men2025, abramovich2025gonetgeneralizabledeeplearning}. In summary, the model was trained on all but one dataset, with the left-out dataset (target domain) used for evaluation, allowing us to assess generalization performance across unseen datasets. To address dataset size imbalances during training, we ensured that the validation set contained a comparable number of images from each of the included training datasets - between a 100 for STARE to 772 for AREDS. The model was trained using a class-weighted cross-entropy loss function, where the weights were computed based on inverse class frequency weighting, ensuring that each class contributed equally to the loss function. Specifically, the weight for each class was set such that $N_0 \cdot w_0 = N_1 \cdot w_1 $ where \( N_c \) and \( w_c \) denote the number of samples and weight for class \( c \), respectively. Other training hyperparameters were similar to those reported in Section \ref{sec:model_benchmark}. We denote AMDNet, the best performing model.

\subsection*{Benchmark to state-of-the-art}
We benchmark our final best performing model, denoted AMDNet, to a SOTA open-source model called DeepSeeNet \cite{deepseenet2019, DeepSeeNet-github}. To focus on our task, i.e., intermediate to late AMD identification, we used both the models LA-Net (late AMD identification) and D-Net (drusen size recognition) provided by DeepSeeNet.

%%%%%% Results plots %%%%%%
\begin{center}
\begin{figure*}[h!]
\vspace{-1em}
    \centering

    \begin{subfigure}{0.85\textwidth}
        \centering
        \begin{minipage}[t]{0.95\textwidth}
            \raggedright
            \textbf{(a)} \label{fig:benchmark}
        \end{minipage}
        \includegraphics[width=\textwidth, trim=0cm 0cm 2cm 7cm, clip]{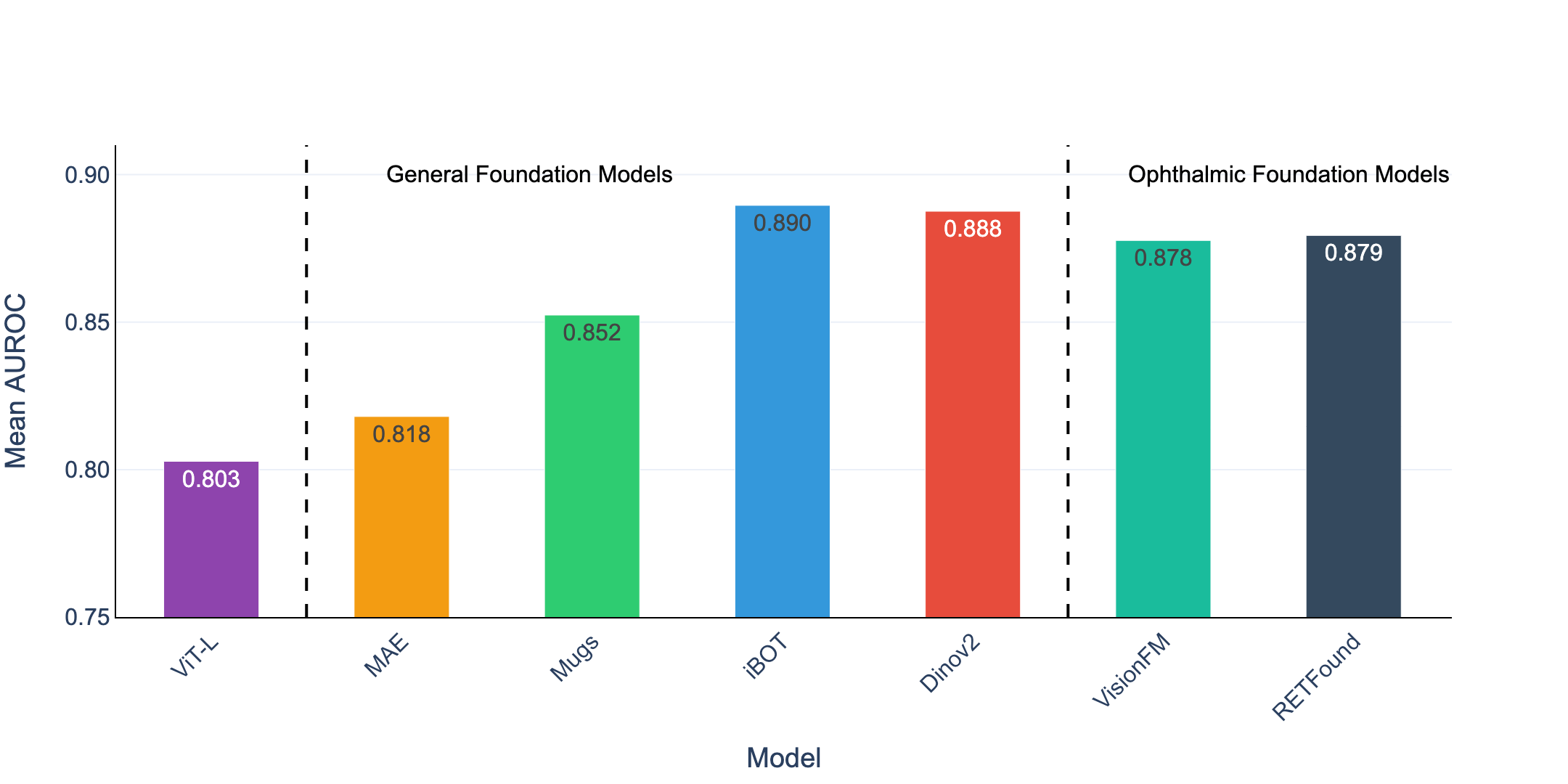}
    \end{subfigure}

    \vspace{-0.8em} % Adjust spacing

    \begin{subfigure}{0.85\textwidth}
        \centering
        \begin{minipage}[t]{0.95\textwidth}
            \raggedright
            \textbf{(b)} \label{fig:amdnet}
        \end{minipage}
        \includegraphics[width=\textwidth, trim=0cm 0.8cm 2cm 4cm, clip]{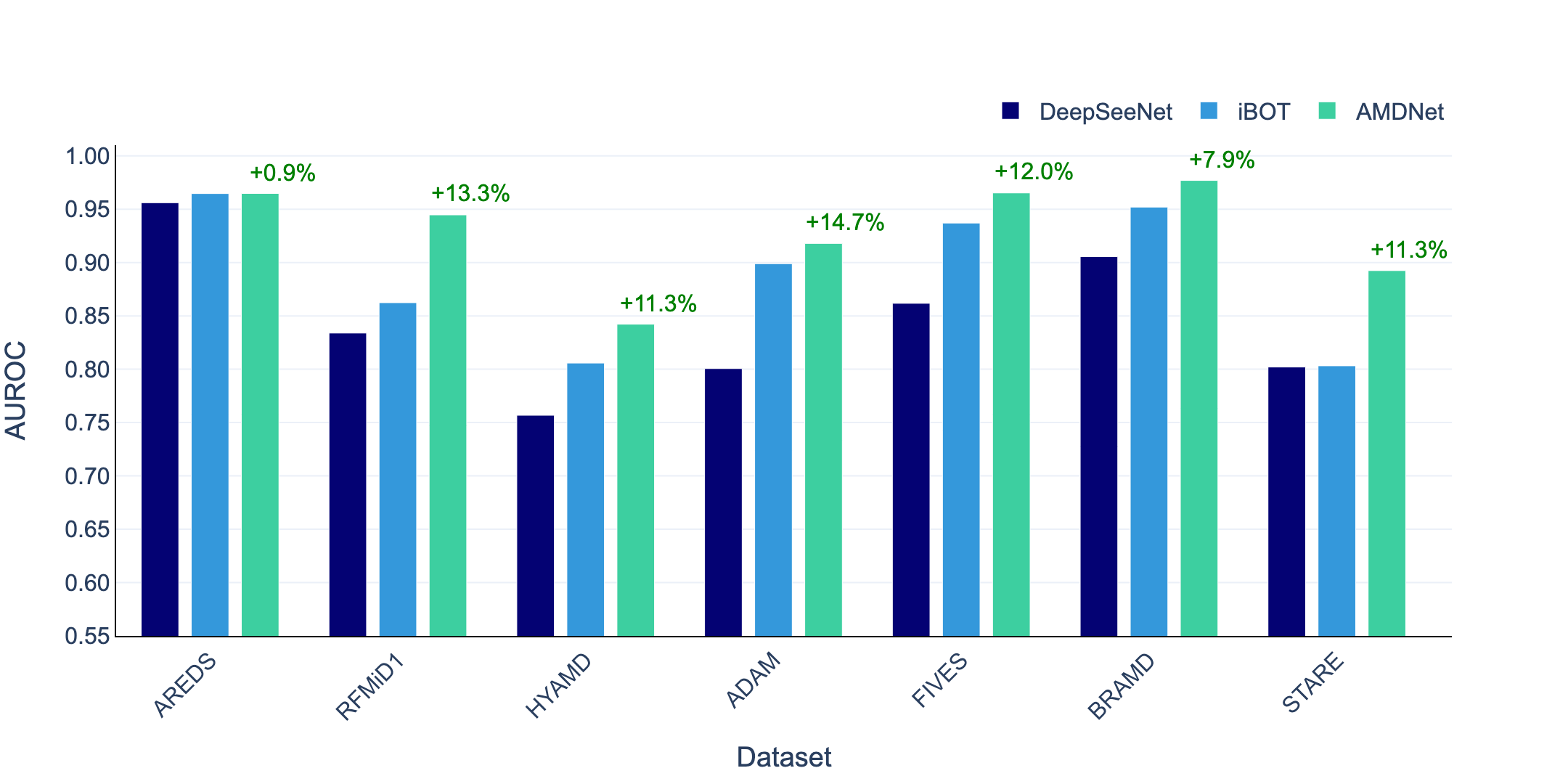}
        \vspace{-0.2em}
    \end{subfigure}

    \vspace{-0.8em}

    \caption{Models performance for AMD identification. (a) Results for a ViT-L model (no pretraining) as well as a set of six foundation models fine-tuned on AREDS-train. The performance is presented as the average AUROC over the AREDS-test and the six external datasets (target domains). (b) The best backbone identified, namely iBOT, is fine-tuned using an all-against-one multi-source domain training approach. For the resulting model, denoted AMDNet, OOD performance is reported for each domain. AMDNet performance is also compared to a state-of-the-art AMD detection model called DeepSeeNet \cite{deepseenet2019}.}
    \label{fig:combined}
\end{figure*}
%%%%%%%%%%%%

%%%%% Results table %%%%%
\begin{table*}
\vspace{-1.5em}
\centering
\caption{AUROC values for seven models. ViT-L refers to a model without pretraining and only trained on AREDS-train. The six following foundation models are fine-tuned solely on the AREDS-train dataset. AMDNet uses an iBOT backbone and employs a multi-source domain (MSD) leave-one-domain-out fine-tuning strategy. The mean and standard deviation are reported for each target domain, obtained by bootstrapping 1,000 samples using 80\% of the dataset. The highest out-of-distribution (OOD) results are marked with \textsuperscript{\textdagger}.}
\label{tab:results}
\renewcommand{\arraystretch}{1.5} % Adjust row height for better readability
\resizebox{\textwidth}{!}{%
\begin{tabular}{lccccccc}
\hline
\textbf{Model} & 
\begin{tabular}[c]{@{}c@{}}\textbf{AREDS} \\\textbf{AUROC}\end{tabular} & 
\begin{tabular}[c]{@{}c@{}}\textbf{RFMiD1} \\\textbf{AUROC}\end{tabular} & 
\begin{tabular}[c]{@{}c@{}}\textbf{HYAMD} \\\textbf{AUROC}\end{tabular} & 
\begin{tabular}[c]{@{}c@{}}\textbf{ADAM} \\\textbf{AUROC}\end{tabular}& 
\begin{tabular}[c]{@{}c@{}}\textbf{FIVES} \\\textbf{AUROC}\end{tabular} & 
\begin{tabular}[c]{@{}c@{}}\textbf{BRAMD} \\\textbf{AUROC}\end{tabular} & 
\begin{tabular}[c]{@{}c@{}}\textbf{STARE} \\\textbf{AUROC}\end{tabular} \\
\hline
%\textbf{Swinv2} & 0.966 $\pm$ 0.002 & 0.833 $\pm$ 0.014 & 0.769 $\pm$ 0.015 & 0.881 $\pm$ 0.015 & 0.905 $\pm$ 0.012 & 0.922 $\pm$ 0.012 & 0.785 $\pm$ 0.019 \\
\textbf{ViT-L} \cite{dosovitskiy2021imageworth16x16words} & 0.913 $\pm$ 0.004 & 0.774 $\pm$ 0.020 & 0.728 $\pm$ 0.017 & 0.761 $\pm$ 0.021 & 0.830 $\pm$ 0.017 & 0.929 $\pm$ 0.012 & 0.682 $\pm$ 0.031 \\
\textbf{MAE} \cite{mae2021} & 0.927 $\pm$ 0.004 & 0.776 $\pm$ 0.019 & 0.735 $\pm$ 0.017 & 0.802 $\pm$ 0.019 & 0.844 $\pm$ 0.015 & 0.939 $\pm$ 0.011 & 0.705 $\pm$ 0.028 \\
\textbf{Mugs} \cite{Mugs2022} & 0.959 $\pm$ 0.003 & 0.833 $\pm$ 0.016 & 0.757 $\pm$ 0.016 & 0.848 $\pm$ 0.018 & 0.886 $\pm$ 0.012 & 0.943 $\pm$ 0.011 & 0.742 $\pm$ 0.023 \\
\textbf{iBOT} \cite{zhou2022ibotimagebertpretraining} & 0.965 $\pm$ 0.002 & 0.862 $\pm$ 0.014 & 0.806 $\pm$ 0.014 & 0.900 $\pm$ 0.014 & 0.937 $\pm$ 0.009 & 0.952 $\pm$ 0.010 & 0.803 $\pm$ 0.023 \\
\textbf{Dinov2} \cite{dinov2-2024} & \textbf{0.969 $\pm$ 0.002} & 0.864 $\pm$ 0.014 & 0.778 $\pm$ 0.015 & 0.900 $\pm$ 0.015 & 0.921 $\pm$ 0.012 & 0.947 $\pm$ 0.010 & 0.834 $\pm$ 0.023 \\
\textbf{RETFound} \cite{retfound2023} & 0.963 $\pm$ 0.003 & 0.860 $\pm$ 0.014 & 0.798 $\pm$ 0.014 & 0.870 $\pm$ 0.017 & 0.922 $\pm$ 0.010 & 0.960 $\pm$ 0.009 & 0.782 $\pm$ 0.027 \\
\textbf{VisionFM} \cite{visionfm2024} & 0.962 $\pm$ 0.003 & 0.872 $\pm$ 0.015 & 0.801 $\pm$ 0.014 & 0.869 $\pm$ 0.017 & 0.916 $\pm$ 0.011 & 0.943 $\pm$ 0.010 & 0.781 $\pm$ 0.021 \\
\hline
\textbf{DeepSeeNet} \cite{deepseenet2019} & 0.956 $\pm$ 0.003 & 0.834 $\pm$ 0.015 & 0.757 $\pm$ 0.015 & 0.801 $\pm$ 0.017 & 0.862 $\pm$ 0.015 & 0.906 $\pm$ 0.014 & 0.802 $\pm$ 0.030 \vspace{0.3em} \\ 
\renewcommand{\arraystretch}{0.9}
\begin{tabular}[c]{@{}l@{}}
\hspace{-0.3em}\textbf{AMDNet} \\
\hspace{1em}{\scriptsize \textit{(iBOT + MSD)}}
\end{tabular} & 0.965 $\pm$ 0.003 & \textbf{0.945 $\pm$ 0.007}\textsuperscript{\textdagger} & \textbf{0.842 $\pm$ 0.012}\textsuperscript{\textdagger} & \textbf{0.918 $\pm$ 0.013}\textsuperscript{\textdagger} & \textbf{0.965 $\pm$ 0.008}\textsuperscript{\textdagger} & \textbf{0.977 $\pm$ 0.007}\textsuperscript{\textdagger} & \textbf{0.892 $\pm$ 0.022}\textsuperscript{\textdagger} \\
\hline
\end{tabular}%
}

\end{table*}

\end{center}
%%%%%%%%%%%%

%%%
\subsection*{Performance Measures}
\label{subsec:performance_measures}
For each target domain, the mean area under the receiver operating characteristic curve (AUROC) is reported along with standard deviation, derived from bootstrapping 1,000 samples, using 80\% of the target domain dataset. When reporting AUROC across all target domains in Figure \ref{fig:combined}a, we provide the median AUROC across the seven domains.  To visualize the decision-making process of AMDNet, we generated heatmaps using Grad-CAM \cite{gradcam}. Specifically, we extracted the final attention layer and weighted it by the gradient between this layer and the output logits.

\subsection*{Error analysis}
The classification performance as a function of the AMD subgroup (LD, GA, NVAMD) was evaluated on the AREDS dataset. To assess the discriminative power of the model for each AMD subgroup, we computed the AUROC by comparing each subgroup against the control group (non-AMD) and while excluding the other two AMD subgroups from the analysis. To assess potential biases in model performance with respect to demographic variables, we evaluated the AUROC separately across patient sex and age groups within the BRAMD dataset. In addition, we stratified the AMD cohort into three clinically meaningful age bins: 18--70, 70--80, and $>85$ years \cite{Jonasson2011}. For each age bin, we computed the AUROC by comparing all control patients (i.e., the entire control cohort) against only the AMD patients within the selected age range.
Furthermore, to examine how comorbidities influence AMD detection, we computed the false positive ratio for each disease in the RFMiD1 dataset, for a threshold of $0.5$. This dataset represents a large variety of diseases, making it a useful tool for error analysis on comorbidities. We then focused our analysis on the six diseases whose false positive ratios exceeded 30\%, thereby highlighting comorbidities most prone to yield misclassifications. %Additionally, we compute a correlation matrix for RFMiD1, to complete the latter analysis.

\section{Results}

%%%
\subsection*{Benchmark of foundation models}
Table \ref{tab:results} summarizes the performance of six foundation models and a baseline ViT-L all fine-tuned on AREDS-train. Figure \ref{fig:combined}a shows the average AUROC performance across all test datasets. Overall, the foundation models outperformed the baseline ViT-L (p<0.05) with an average performance improvement of 8\%. Among the six foundation models benchmarked, iBOT achieved an average AUROC of 0.890 across all test datasets, slightly outperforming Dinov2 which obtained 0.888. Additionally, iBOT demonstrated superior OOD generalization in three out of six target domains: HYAMD, ADAM, and FIVES. When comparing general foundation models pre-trained on natural images to those pre-trained in-domain, the results show that, in five out of six target domains, the general foundation models performed better. Only in the case of BRAMD did VisionFM outperform the general foundation models.

%While Dinov2 \cite{dinov2-2024} outperformed iBOT \cite{zhou2022ibotimagebertpretraining} on AREDS-test with an AUROC of 0.969 versus 0.965,

%%%
\subsection*{AMDNet}
The best foundation model, denoted AMDNet, consists of iBOT fine-tuned using the multi-source approach. When comparing AMDNet to iBOT fine-tuned on the single-source domain AREDS-train, we found that AMDNet improved performance by 4.7\%. When compared to an open-source SOTA AMD identification algorithm called DeepSeeNet \cite{deepseenet2019}, AMDNet demonstrated significantly superior and non-incremental performance improvements. On average, AMDNet achieved an AUROC 10.2\% higher than DeepSeeNet.

%\subsection*{Ablation Study}
%TODO.

%%
\subsection*{Error Analysis}
\begin{figure*}
    \centering

    \begin{subfigure}[b]{0.8\textwidth}
        \centering
        \begin{minipage}[t]{0.95\textwidth}
            \raggedright
            \textbf{(a)}
        \end{minipage}
        \includegraphics[width=\textwidth]{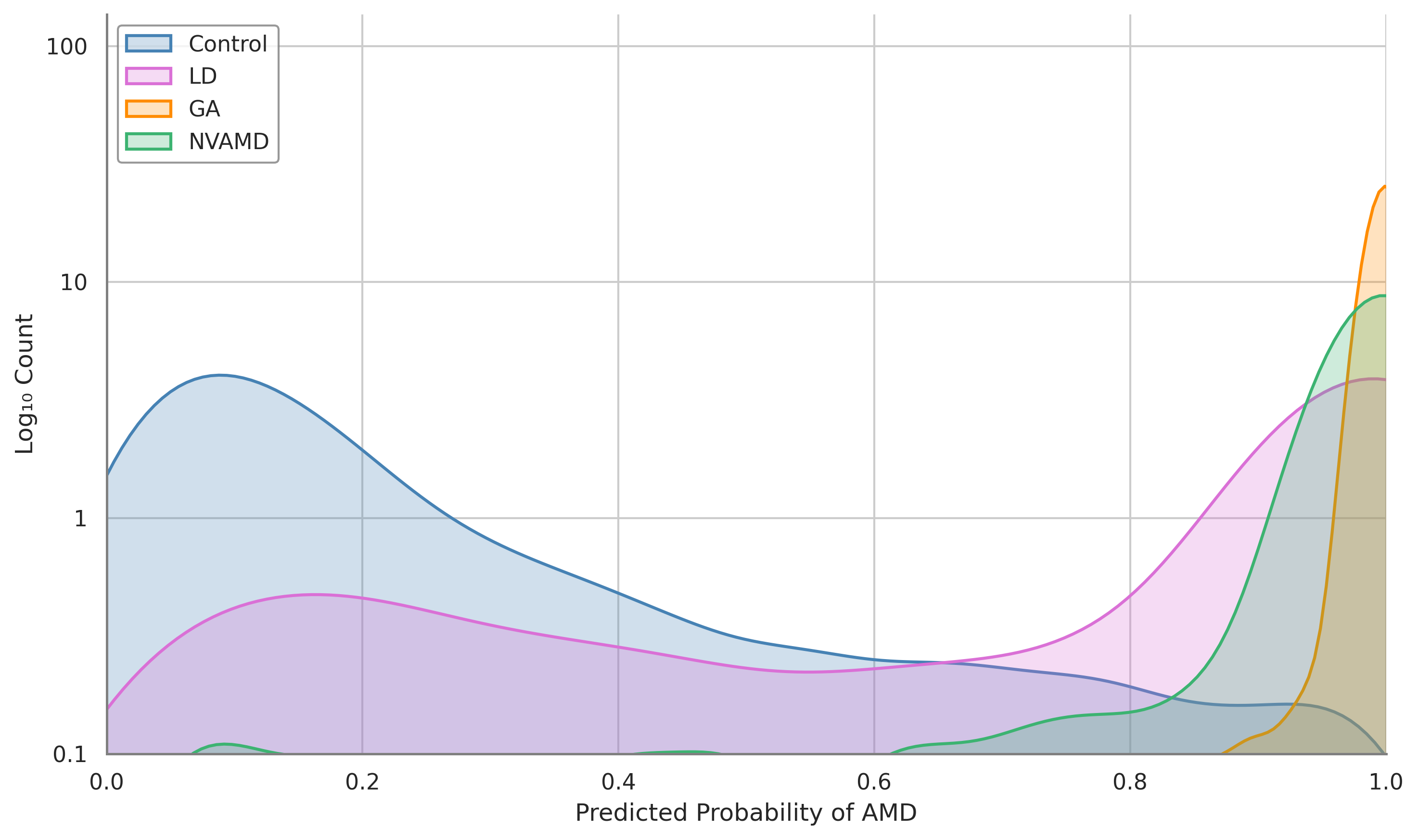}
    \end{subfigure}

    \vspace{1em}

    \begin{subfigure}[b]{0.45\textwidth}
        \centering
        \begin{minipage}[b]{0.95\textwidth}
            \raggedright
            \textbf{(b)} 
        \end{minipage}
        \includegraphics[width=\textwidth]{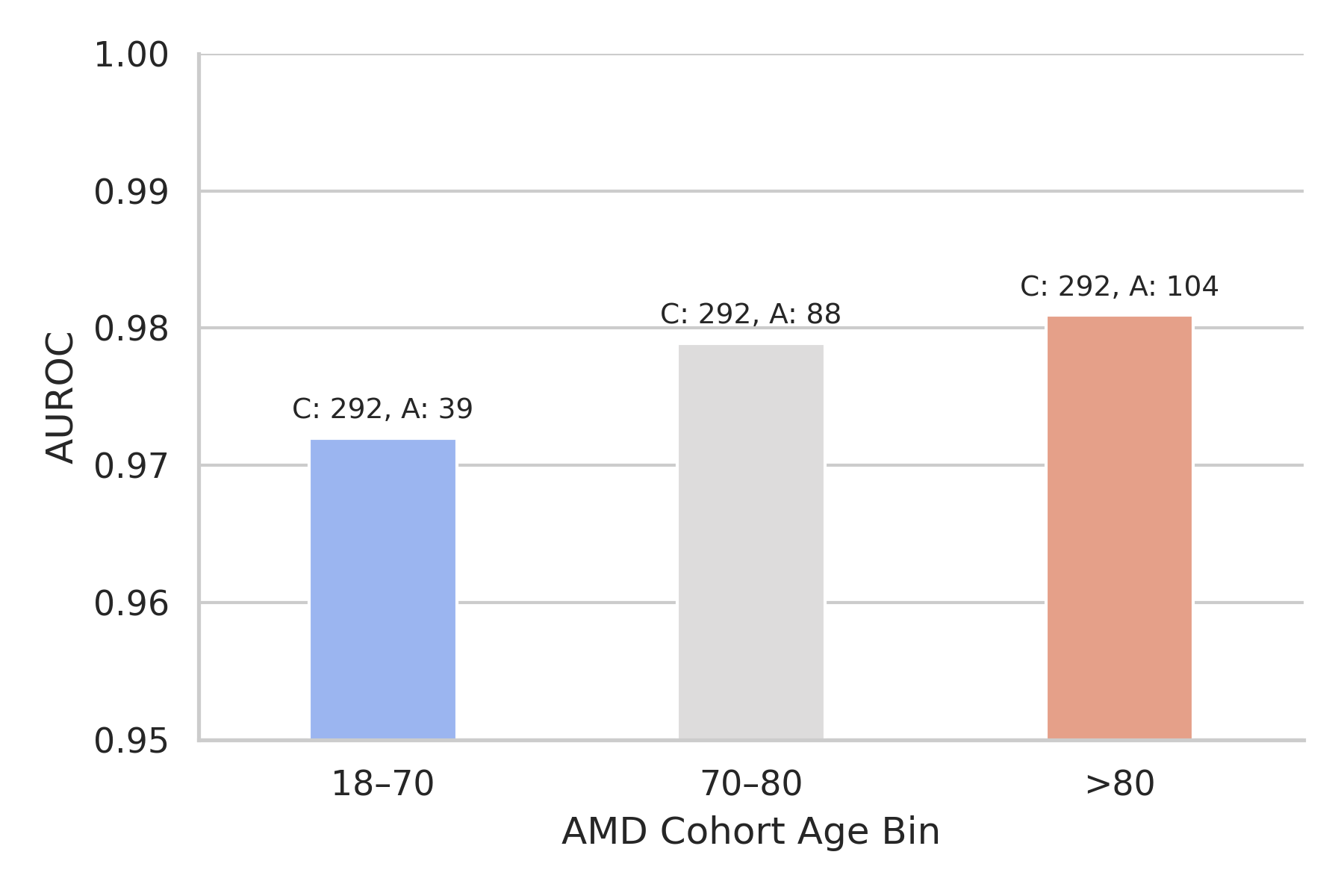}
    \end{subfigure}
    \hfill
    \begin{subfigure}[b]{0.45\textwidth}
        \centering
        \begin{minipage}[b]{0.95\textwidth}
            \raggedright
            \textbf{(c)} 
        \end{minipage}
        \includegraphics[width=\textwidth]{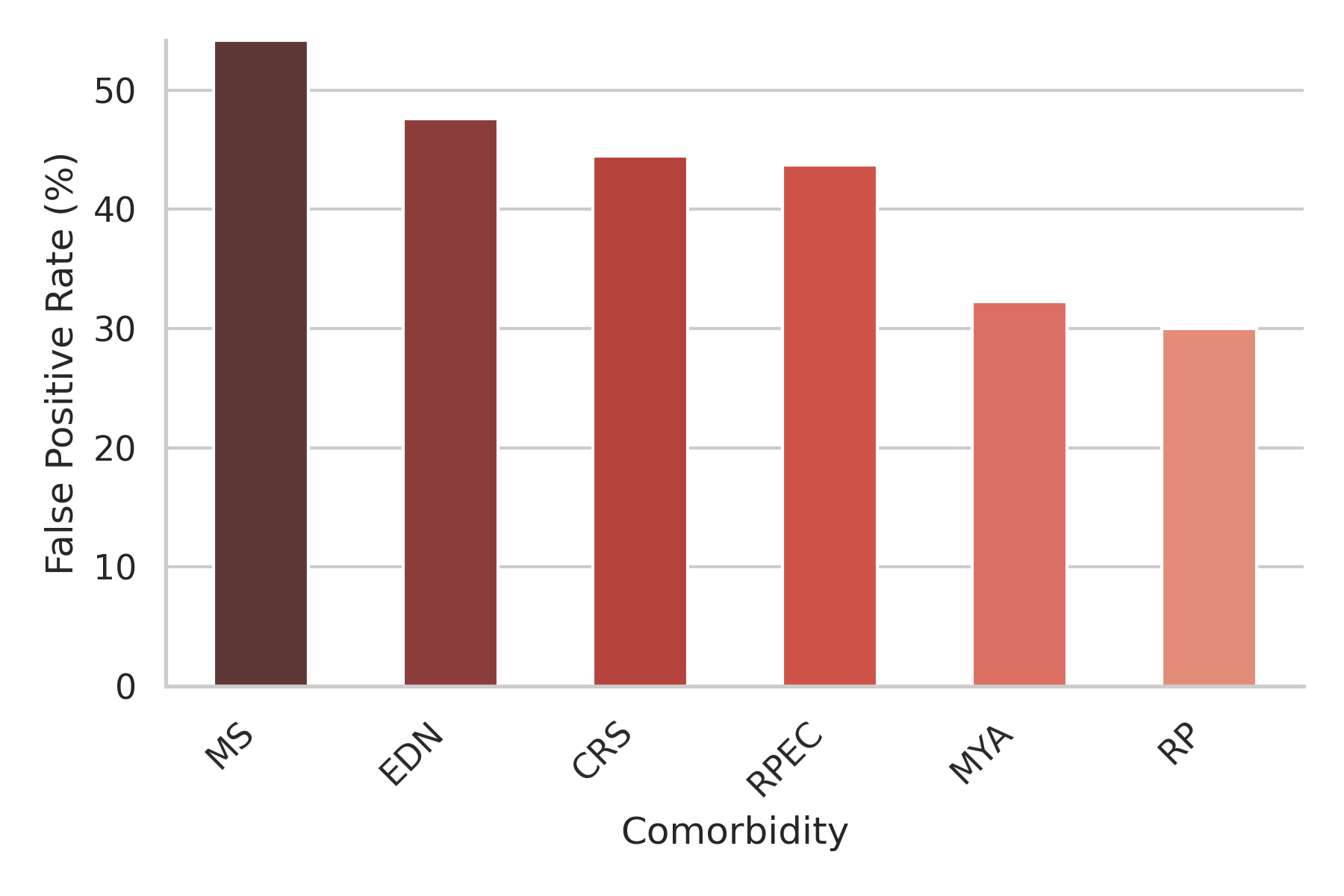}
    \end{subfigure}

    \vspace{-0.3em}

    \caption{Error analysis. a) AMDNet probability output for the different subgroups for AREDS; b) AUROC per age group for BRAMD. C: number of control images, A: number of AMD images in the age bin; c) False positive rate per comorbidity for RFMiD1. LD: large drusen; GA: geographic atrophy; NVAMD: Neovascular AMD.}
    \label{fig:analysis_plots}
\end{figure*}

The model achieved an AUROC of 0.930 for LD versus 0.998 for GA and 0.980 for NVAMD, indicating more difficulty in identifying LD than the other AMD subgroups. AUROC scores were consistently high for both sexes, with 0.975 for male patients and 0.979 for female patients, indicating an absence of significant sex-based differences in classification outcomes. For age, the model achieved its highest AUROC for the $>85$ age group (0.981). Performance was slightly lower in the younger groups, with AUROC values of 0.972 and 0.979 respectively for the 18--70 and 70-80 groups (Figure \ref{fig:analysis_plots}b).
We found the most prevalent comorbidities among false positives (FP) are Macular Scar (MS), Exudation (EDN), Chorioretinitis (CRS), Retinal pigment epithelium changes (RPEC), Myopia (MYA), and retinitis pigmentosa (RP) with a FP rate of, respectively, 54.2\%, 48.6\%, 44.4\%, 43.8\%, 42.3\%  and 30.0\%. Other conditions show a FP rate lower than 30\% for the given threshold. A correlation matrix reveals that only MYA is correlated to AMD, with a factor of 0.2. Results of the comorbidity analysis are shown in Figure \ref{fig:analysis_plots}c.

%%%%%% Statistics figures %%%%%%
%%%%%%%%%%%%

%%%%%% Heatmaps figure %%%%%%
\begin{figure*}[t]
    \centering
    % First row
    \begin{subfigure}[b]{0.48\textwidth} % Adjust width as needed
        \centering
        \includegraphics[width=\textwidth]{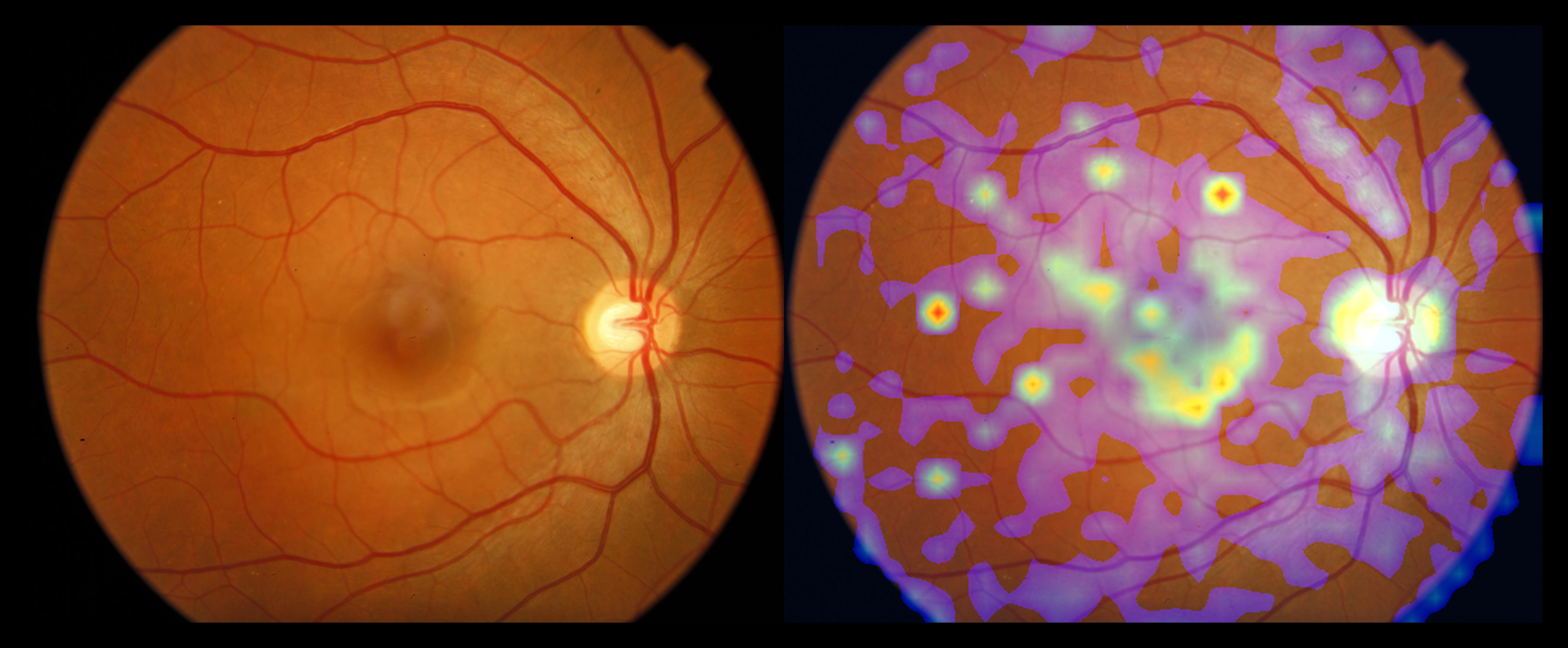}        \caption*{(a) Retina without AMD. $p = 0.04$}
        \label{fig:image1}
    \end{subfigure}
    \hspace{15pt}% Horizontal space between subfigures
    \begin{subfigure}[b]{0.48\textwidth} % Adjust width as needed
        \centering
        \includegraphics[width=\textwidth]{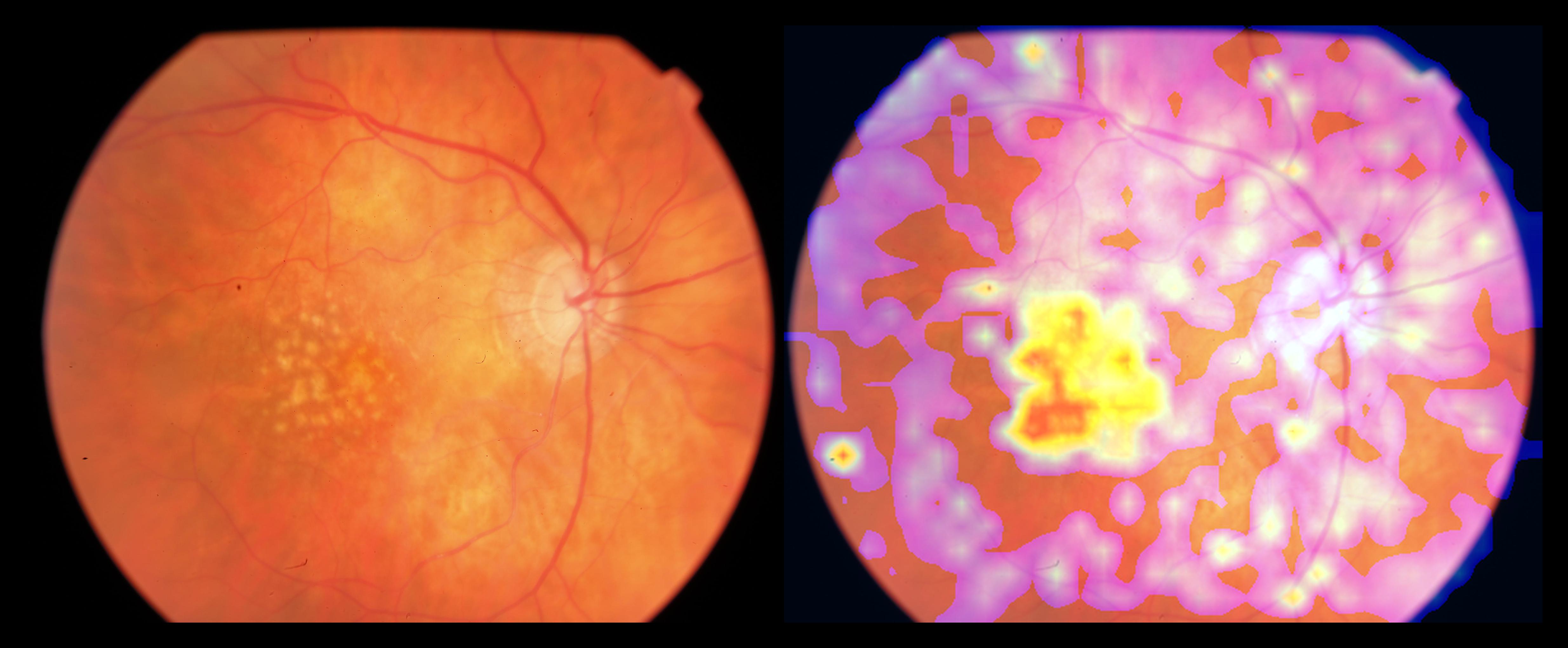}
        \caption*{(b) Retina with Large Drusen. $p = 0.99$}
        \label{fig:image2}
    \end{subfigure}

    % Space between rows
    \vspace{8pt} 

    % Second row
    \begin{subfigure}[b]{0.48\textwidth} % Adjust width as needed
        \centering
        \includegraphics[width=\textwidth]{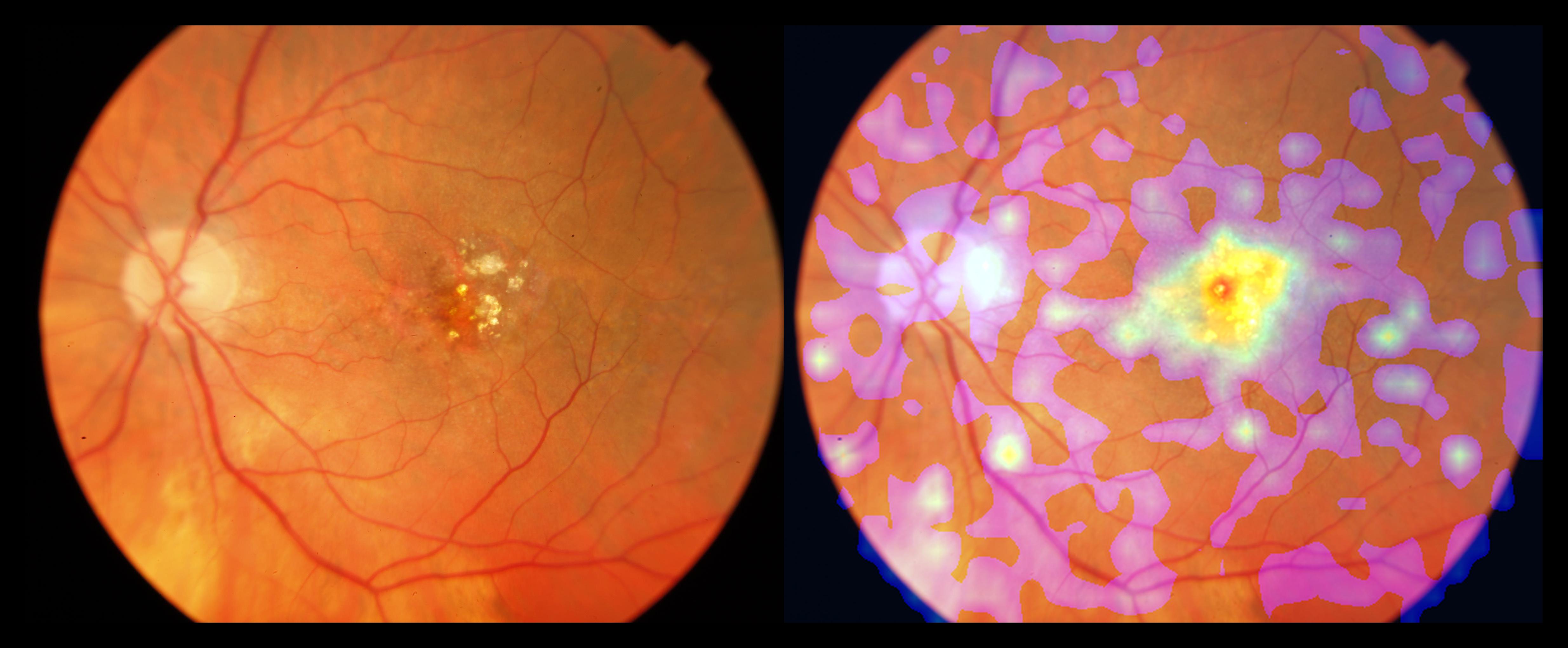}
        \caption*{(c) Retina with GA. $p = 0.96$}
        \label{fig:image3}
    \end{subfigure}
    \hspace{15pt}% Horizontal space between subfigures
    \begin{subfigure}[b]{0.48\textwidth} % Adjust width as needed
        \centering
        \includegraphics[width=\textwidth]{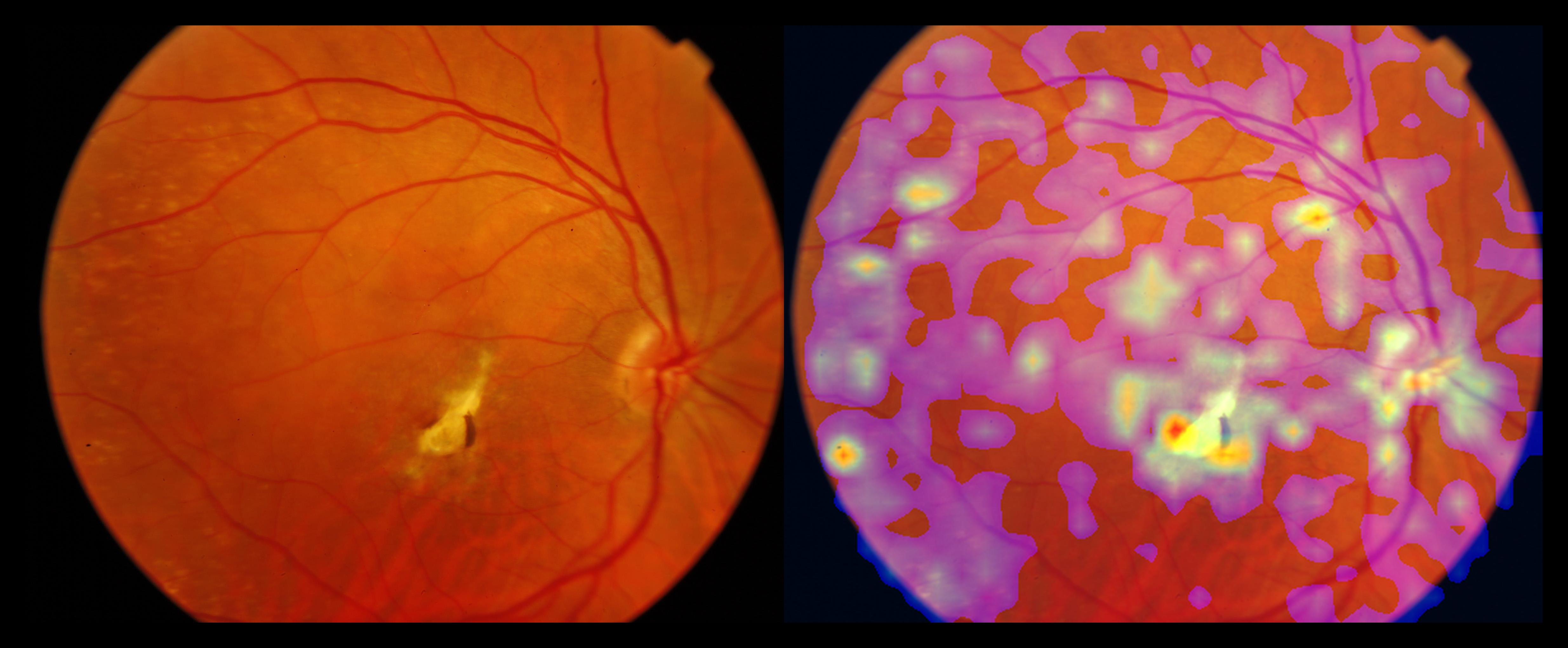}
        \caption*{(d) Retina with NVAMD. $p = 0.68$}
        \label{fig:image4}
    \end{subfigure}

    \caption{Examples of attention maps of AMDNet on images from BRAMD. We used GradCAM \cite{gradcam} to plot in red the regions used by the model to generate its predictions. $p$ is the probability for AMD given by AMDNet.}
    \label{fig:attention_maps}
\end{figure*}
%%%%%%%%%%%%

\section*{Discussion}
%%%%%%%%%%
% Need to write a discussion discussing foundation models and relating to our findings.
In this research, we evaluated six computer vision foundation models for the task of intermediate to late AMD identification. On average, the iBOT backbone exhibited superior OOD  performance (Figure \ref{fig:combined}a and Table \ref{tab:results}) although it was in pair with DinoV2. Notably, the performance of RETFound and VisionFM, two recent DFI-based foundation models, were inferior to iBOT, which is a foundation model pretrained using natural images. This suggests that the use of a large number of in-domain images, as in RETFound or VisionFM, do not provide an advantage over using a very large number of natural images for pretraining with SSL for the task of moderate to late AMD identification. This finding confirms our earlier, albeit less thorough observations, where we observed that a DinoV2 backbone was found to yield better outcomes for diabetic retinopathy staging \cite{Men2025} and glaucoma identification \cite{abramovich2025gonetgeneralizabledeeplearning} from DFI compared to a RETFound backbone. In addition, a recent study by Xiong et al. \cite{xiong2025} evaluated the performance of RETFound compared to a pre-trained ViT on ImageNet using an Asian population sample for glaucoma and coronary heart disease diagnosis, as well as 3-year risk prediction. They found no statistically significant improvement. Taken together, our findings challenge the value of recent contributions in developing in-domain retinal foundation models.

As part of this research, we developed a new SOTA model for the identification of moderate to late AMD: AMDNet. To do so, we used the iBOT backbone and a multi-source training approach. We had shown the value of this training approach in improving OOD generalization \cite{Men2025, abramovich2025gonetgeneralizabledeeplearning}. AMDNet has 0.842-0.977 OOD AUROC and outperformed an open source SOTA model DeepSeeNet. As part of the error analysis, we found that AMDNet performed best in detecting advanced AMD subgroups (GA and NVAMD) versus the LD subgroup (Figure \ref{fig:analysis_plots}a). We also found that AMDNet performed better at identifying GA versus NVAMD. Given the fact NVAMD is characterized by the appearance of new vasculature and accumulation of subretinal fluid, which can be of small size and hard to differentiate from the normal vasculature of the eye, this stage is harder to diagnose than GA which is a visible atrophy of the retina, showing well demarcated, brighter and abnormal lesions on the macula (Figures \ref{fig:attention_maps}c and \ref{fig:attention_maps}d). The error analysis also showed that false positive cases were associated with comorbidities affecting the retina in a manner that can mimic AMD: targeting the macula (eg. MS), changes in pigmentation (eg. RPEC, RP), apparition of drusen or similar lesions (eg. EDN, CRS). Analysis of the model's performance across different age groups revealed a slight trend of increasing accuracy with patient age (Figure \ref{fig:analysis_plots}b). This finding is likely explained by the age-related nature of AMD: older patients often exhibit more advanced disease stages, which are typically easier for the model to detect.

% , etc. MYA and AMD seem to be correlated, as the correlation factor of 0.2 indicates, and that would be an explanation for the relatively high FP rate among MYA patients

When investigating AMDNet using explainability maps, we observed that the model consistently focuses its attention on the macula, even in healthy images (Figure \ref{fig:attention_maps}a). AMDNet accurately detects lesions such as LD (Figure \ref{fig:attention_maps}b), GA (Figure \ref{fig:attention_maps}c), and NVAMD (Figure \ref{fig:attention_maps}d). However, in the case of neovascular lesions in Figure \ref{fig:attention_maps}d, although the vasculature was correctly localized, the model assigned only a moderate probability (0.68) for the image being classified in the AMD cohort. This moderate confidence could be attributed to the lesion's location, slightly outside the macular region, or the absence of additional characteristic lesions such as drusen.

The detection of AMD in DFIs has gained increasing attention, driven by advances in imaging technology and computational analysis \cite{Pead2019}. DFIs enable high-resolution visualization of retinal structures, facilitating the detection of drusen deposits, pigmentary abnormalities, and neovascular changes characteristic of AMD \cite{Ferris2013}. Advanced machine learning and deep learning algorithms have been developed to automate the analysis of these images, enhancing diagnostic accuracy and consistency \cite{Grassmann2018, deepseenet2019}. These automated systems mitigate inter-observer variability inherent in manual interpretations and allow for early detection, which is crucial for timely intervention and management of AMD progression \cite{Burlina2017}. Moreover, integrating DFIs with telemedicine platforms has expanded accessibility to AMD screening, particularly in under-served regions lacking specialized ophthalmic services \cite{Wang2025}. The scalability of DFIs combined with automated analysis supports large-scale screening programs, facilitating population-wide assessments, treatments, and monitoring. The majority of prior works \cite{Mookiah2014,deepseenet2019, Zapata2020, Xie2023} have focused on single-population datasets, which restricts evaluating how these models performs across diverse populations, retinal cameras and clinical settings. Recent research has evaluated performances  of these models on external datasets, revealing significant performance degradation \cite{Grassmann2018, amd-progression}. By leveraging foundation models and a multi-source domain training approach, this research showed the feasibility of training a deep learning for intermediate-to-late AMD identification with consistently high OOD performances.

This research has several limitations. Although the study includes a relatively large number of datasets, the model should be tested on additional datasets from medical centers worldwide to encompass a broader range of ethnicities, comorbidities, medical practices, and camera types and FOVs. These datasets should better reflect real-life scenarios, as most publicly available datasets are inherently biased due to their creation process, and therefore do not represent the intended population. One of the main limitation in our final AMDNet model is the limitations in its explainability. Indeed, AMDNet does not indicate and segment specific objects (e.g., drusen, GA) which may be useful to the clinical decision making process. Finally, although we found that RETFound and VisionFM provided no added value on our downstream task, further benchmarks on various ophthalmic and systematic downstream task are wanted in order to further confirm our findings.

Overall, this research explored the effectiveness of foundation models for intermediate-to-late AMD identification from single DFI images. Our quantitative results show that general foundation models outperformed domain-specific models for this task. The best-performing model, AMDNet, demonstrates robustness and achieved high OOD generalization performance. Additionally, this work contributes to the field by providing BRAMD, a new dataset we developed, as an open-access resource. AMDNet will be accessible at [URL upon publication].

\section{Acknowledgments}

The Age-Related Eye Disease Study (AREDS) Database can be found at \href{https://www.ncbi.nlm.nih.gov/projects/gap/cgi-bin/study.cgi?study_id=phs000001.v3.p1}{this URL} through dbGaP accession number phs000001.v3.p1. Funding support for AREDS was provided by the National Eye Institute (N01-EY-0-2127). We would like to thank the AREDS participants and the AREDS Research Group for their valuable contribution to this research. This study/research project is/was (partially) supported by the Technion EVPR Fund: Irving \& Branna Sisenwein Research Fund.

% Use the style of numbering in square brackets.
% If nothing is used, default style will be taken.
%\begin{enumerate}[a)]
%\item 
%\item 
%\item 
%\end{enumerate}  

%\begin{itemize}
%\item 
%\item 
%\item 
%\end{itemize}  

%\begin{description}
%\item[]
%\item[] 
%\item[] 
%\end{description}  

\printcredits

%\bibliographystyle{elsarticle-num}

\bibliographystyle{elsarticle-num}
\bibliography{references}



\end{document}